# Exploring novel prognostic biomarkers and biologic processes involved in NASH, cirrhosis and HCC based on survival analysis using systems biology approach

Dr. Sedigheh Behrouzifar

Department of Medical Sciences, Shahrood Branch,

Islamic Azad University, Shahrood, Iran

sedighehbehrouzifar@gmail.com

**Abstract**

*There is an unmet need to develop medications or drug combinations which can stop advancement of NASH to liver cirrhosis and HCC. Therefore, identifying key biomarkers based on overall survival and the exploring biological processes involved in the pathogenesis and progression of NASH toward cirrhosis and HCC to improve therapeutic interventions is necessary. The microarray dataset from the GPL13667 platform was downloaded from the Gene Expression Omnibus (GEO). The inclusion criteria for the DEGs included an adjusted p-value <0.05 and a log(2) fold change >1. In the first step, protein-protein interaction (PPI) network of differentially expressed genes (DEGs) in every three groups (NASH, Cirrhosis and HCC) was constructed using STRING online database. In the second step, the DEGs of each group were imported to Cytoscape software separately, and the genes with Degree $\geq 3$ were selected. In the third step, the genes with Degree $\geq 3$ were imported to Gephi software (0.9.2 version) and the genes with betweenness centrality >0 were selected. Venn diagram of these genes was depicted for the NASH, cirrhotic and HCC groups. According to venn diagram, 96 (NASH), 30 (cirrhotic) and 213 (HCC) genes were specifically upregulated (based on inclusion criteria). Among specific genes, 22 (NASH), 5 (cirrhotic) and 82 (HCC) genes were with poor overall survival. The overlap among the 3 groups (NASH, Cirrhosis and HCC) contained 4 upregulated genes HLA-F, HLA-DPA1, TPM1 and YWHAZ. From 4 upregulated genes, only YWHAZ gene was with poor overall survival. Probably, upregulated genes with poor overall survival in NASH and cirrhosis are the biomarkers and predictors for infecting NASH and Cirrhotic patients to HCC in future. In three groups (NASH, cirrhosis and HCC), YWHAZ gene was enriched in several biological processes including cytokine-mediated signaling pathway, negative regulation of programmed cell death, regulation of MAPK cascade, regulation of mRNA stability and cellular component assembly. The present study detected new candidate genes and key biological processes in NASH, cirrhosis and HCC based on overall survival and using in silico analysis. Therefore, performing in vitro and in vivo researches to verify the results is necessary.*

## 1. Introduction:

Nonalcoholic fatty liver disease(NAFLD) has appeared as the most important cause of chronic liver disease in many world regions (Del Ben et al., 2014). NAFLD affects about quadrant of the global population (Younossi et al., 2016). As diabetes and metabolic syndrome is increasing, the prevalence of NAFLD is also developing (Friedman et al., 2018). NAFLD occurs usually in overweight or obese people (Francque et al., 2021). World Health Organization estimates that the global percentage of overweight or obese adults will increase 57.8% by 2030 (Streba et al., 2015). However, the prevalence rate of lean/non-obese NAFLD varies between 3% and 30% worldwide (Kuchay et al., 2021). NAFLD can progress from simple steatosis to non-alcoholic steatohepatitis (NASH), cirrhosis and even hepatocellular carcinoma (HCC) (Asrih and Jornayvaz, 2015, Wong et al., 2015, Sheka et al., 2020). Insulin resistance, obesity, dyslipidemia, and altered lipid metabolism synergistically accelerate the accumulation of hepatic triglycerides, lipotoxicity, and inflammation that are necessary for progression of NAFLD toward NASH and cirrhosis (Farrell et al., 2018). Pathogenesis of NAFLD is characterized by releasing pro-inflammatory cytokines, as well as downregulation of anti-inflammatory adipokines (Stine et al., 2018). Hypertension is a prevalent comorbidity in lean and overweight/obese-NAFLD (Kuchay et al., 2021). Insulin resistance leads to enormous FFA influx from the adipose tissue to hepatocytes. The liver has a limited capacity to consume, store, produce and transport triglycerides out of the liver cell (Rada et al., 2020). Fatty acid oxidation yields free radicals, resulting in oxidative stress. Reactive oxygen species (ROS) promote the abnormal release of pro-inflammatory cytokines such as tumor necrosis factor-alpha (TNF-α), interleukin-6 (IL-6) and chemokines by hepatocytes and recruit circulating white blood cells (Rada et al., 2020, Dongiovanni et al., 2021). Inflammation and oxidative stress serve the critical roles in the progression of NASH (Zhang et al., 2021b). NASH is characterized by elevated liver enzymes, without a history of alcohol consumption (Satiya et al., 2021). Because NASH is diagnosed by liver biopsy, the prevalence of NASH is underestimated (Ajmera and Loomba, 2021). Non-specific symptoms of NASH include tiredness and developing pain in the upper right quadrant of the abdomen. The lack of specific clinical manifestations in early stages of disease leads to advance hepatocyte damage and poor prognosis (Povsic et al., 2019). Cholesterol overload in hepatocytes may also stimulate hepatic stellate cells (HSCs) activation (Dongiovanni et al., 2021). Moreover, transforming growth factor-β (TGFβ) activates HSCs and is released by several hepatic cells (Schwabe et al., 2020). HSCs are responsible for developing cirrhotic NASH liver (Schwabe et al., 2020). One fifth patients with NASH will progress cirrhosis requiring liver transplant (Sheka et al., 2020). NASH-related HCC may develop in the presence and absence of cirrhosis (Pais et al., 2017). In comparison with non-cirrhotic NASH, cirrhotic NASH patients are in higher risk of HCC (Cholankeril et al., 2017). HCC is third major cause of cancer-associated death in world (Désert et al., 2018). Metastasis is the most important impediment to restore improvement in HCC patients (Chen et al., 2021). Metabolic, immunity, dietary, and genetic-related factors predispose NASH patient to liver cancer (Dongiovanni et al., 2021). In the last 20 years, the number of non-viral-HCC patients has been rapidly rising (Kim et al., 2021). Despite extensive efforts to develop the effectiveness of NASH, cirrhosis and HCC management and enhancement of quality of life, no significant improvement has been made in the disease process yet. Exploring the mechanisms underlying the progression of HCC in the context of NASH and achieving medications or drug combinations that can stop advancement of NASH toward liver cirrhosis and HCC would be required. Therefore, identifying and screening key target proteins based on overall survival as well as insight into the biological processes involved in the pathogenesis and progression of disease in order to improve response to treatment is essential.

## 2. Materials and Methods

The microarray dataset from the GPL13667 platform ([HG-U219] Affymetrix Human Genome U219 Array) was downloaded from the GEO (http://www.ncbi.nlm.nih.gov/geo/). The GSE164760 expression profile contains 6 healthy samples, 8 cirrhotic samples,74 NASH samples and 53 HCC samples.

### 2.1. Inclusion Criteria of the DEGs

For data analysis, R studio software was used. The DEGs between HCC and healthy liver tissues, cirrhotic and healthy liver tissues, NASH and healthy liver tissues were determined. The inclusion criteria for the DEGs included an adjusted p-value <0.05 and a log(2) fold change (logFC) > 1. In the first step, PPI network of DEGs in every three groups (NASH, Cirrhosis and HCC) was constructed using Search Tool for the Retrieval of Interacting Genes (STRING; http://string-db.org) online database. In the second step, the DEGs of each group were imported to Cytoscape software separately, and the genes with Degree $\geq$ 3 were selected. In the third step, the genes with Degree $\geq$ 3 were imported to Gephi software (0.9.2 version) and the genes with betweenness centrality >0 were selected. Venn diagram of the genes identified by Gephi software was depicted for the NASH, cirrhotic and HCC groups using online tool (https://bioinformatics.psb. ugent.be/webtools/Venn/).

Then, the overall survival (OS) of all genes identified by Gephi software was analyzed using Kaplan-Meier curve in GEPIA (Gene Expression Profiling Interactive Analysis). GEPIA website(http://gepia.cancer-pku.cn/) predicts survival biomarkers and analyzes the overall survival of patients with high and low expression of certain genes. Log-rank p<0.05 was considered as poor survival.

### 2.2. Annotation of the DEGs

In this study, the Enrichr (https://maayanlab.cloud/Enrichr/) was used to perform Gene Ontology (GO) enrichment analyses. Through Gene Ontology (GO) enrichment analysis, we can understand DEGs biological Characteristics. The genes identified by Gephi software were imported to Enrich web server. Enrichr analysis was performed for HCC, NASH and Cirrhosis, separately and biological processes of genes with poor overall survival was determined and p-value < 0.05 was considered statistically significant.

### 3. Results
### 3.1. Identification of the DEGs

In NASH, cirrhotic and HCC groups, 413,196 and 576 genes with adjusted p-value<0.05 and log FC>1 were detected respectively. The DEGs in every group were imported to STRING database and PPI was obtained. Then, using Cytoscape software, analysis of network was performed, and the genes with degree$\geq$3 in every group were selected. In NASH, cirrhotic and HCC groups, 199, 50 and 310 genes with degree$\geq$3 were detected respectively. The genes with degree$\geq$3 in every group, were imported to Gephi software, and 187 (NASH), 39 (cirrhotic) and 301 (HCC) genes with betweenness>0 were selected. Then, venn diagram of NASH and cirrhotic and HCC groups was depicted (Figure1). According to venn diagram, 96, 30 and 213 genes were specifically upregulated (based on inclusion criteria) in NASH, cirrhosis and HCC, respectively. From 96 specific genes for NASH, 22 genes were with poor overall survival (suplementary4). From 30 specific genes for Cirrhosis, 5 genes ENG, RHOQ, CCR7, SH2D1A and HDAC7 were with poor overall survival. From 213 specific genes for HCC, 82 genes were with poor overall survival.

According to inclusion criteria, eighty-three shared genes were detected between NASH and HCC (Figure1). From 83 genes, 31 genes were with poor overall survival.

The overlap among the 3 groups (NASH, Cirrhosis and HCC) contained 4 upregulated genes HLA-F, HLA-DPA1, TPM1 and YWHAZ as shown in the Venn diagram (Figure1). Among these four upregulated genes, YWHAZ gene was with poor overall survival. Probably, upregulated genes with poor overall survival in NASH and cirrhosis are the biomarkers and predictors for infecting NASH and cirrhotic patients to HCC in future. Four shared upregulated genes between cirrhotic and NASH groups were including TRIM22, CASP1, IKZF1 and FYB. From 4 upregulated genes, no genes were with poor overall survival. Moreover, the gene STAT1 as only shared gene between Cirrhosis and HCC was not with poor overall survival.

### 3.2. Gene Ontology Enrichment Analysis of the DEGs

Analysis of GO of shared and specific DEGs was performed using Enrichr web server and p-value<0.05 was considered statistically significant. The first, 187 (NASH), 39 (cirrhotic) and 301(HCC) genes with betweenness>0 were imported to Enrichr web server, separately. Then, biological processes enriched by the genes with poor overall survival in three groups (NASH, Cirrhosis, HCC) were explored. For shared gene YWHAZ, biological processes were evaluated in NASH, cirrhosis and HCC separately. In three groups (NASH, cirrhosis and HCC), YWHAZ gene was enriched in biological processes including cytokine-mediated signaling pathway, negative regulation of programmed cell death, regulation of MAPK cascade, regulation of mRNA stability and cellular component assembly (Table1).

Regarding shared prognostic genes (OS<0.05) between HCC and NASH, gene-biological process network was depicted (Figure 2). RAC1 (outdegree=15), HNRNPU(outdegree=13), ITGAV (outdegree=12) and UBE2N (outdegree=10) were with the most outdegree. The shared biological processes with the most indegree included transcription (indegree=5), mRNA processing (indegree=5), protein ubiquitination (indegree=5), neutrophil degranulation (indegree=4) and nucleic acid metabolism (indegree=4). The biological processes with the most indegree were considered as key biological processes. Regarding specific prognostic genes in HCC (OS<0.05), gene-biological process network was depicted (Figure 3). Network analysis of Cytoscape related to specific genes in HCC (OS<0.05) showed that the most outdegree was related to the genes ITGB1 (outdegree=24), PRKCI (outdegree=18), PPIA (outdegree=17) and RAB7A (outdegree=16) respectively. It suggests that these genes may play important roles in the progression of HCC. The most indegree was related to protein modification (indegree=15), transcription (indegree=15) and regulation of apoptosis (indegree=10). Regarding specific prognostic genes in NASH (OS<0.05), gene-biological process network was depicted (Figure 4). Network analysis of Cytoscape related to specific genes of NASH (OS<0.05) showed the most outdegree was related to the genes GSK3B (outdegree=31), DAB2 (outdegree=14), PSME4 (outdegree=14) and PPP3CA (outdegree=9) respectively. The most indegree was related to protein modification (indegree=7), ubiquitination (indegree=6), Wnt signaling (indegree=5) and transcription (indegree=4). Regarding specific genes of cirrhosis (OS<0.05), gene-biological process network was depicted (Figure5). Network analysis of Cytoscape related to specific genes in cirrhosis(OS<0.05) showed that the most outdegree was related to the genes CCR7 (outdegree=34), ENG (outdegree=11), RHOQ (outdegree=7) and HDAC7 (outdegree=6). The most indegree was related to apoptosis (indegree=3) and signal transduction (indegree=3).

### 4. Discussion

According to the present study, YWHAZ gene was highlighted as a key prognostic biomarker in NASH, cirrhosis and HCC. Growing researches have reported that YWHAZ is up-regulated in multiple types of cancers, acting as an oncogene (Yu et al., 2013, Zhao et al., 2018, Xia et al., 2019, Gan et al., 2020). In some studies, in order to suppress cell proliferation and metastasis (invasion and migration), YWHAZ gene was targeted (Zhao et al., 2018, Wei et al., 2019, Jiang et al., 2018, Chen et al., 2016). One study showed that YWHAZ gene is a stably expressed housekeeping gene in Non-alcoholic fatty liver disease (Bruce et al., 2012). However, Hye Ri Ahn et al. (Ahn et al., 2021) revealed that in chronic hepatitis, cirrhosis and HCC, expression level of YWHAZ gene was not stable. In the study of Meile Mo et al. (Mo et al., 2022) the YWHAZ expression level was significantly associated with the infiltrating levels of B cells, neutrophils, macrophages, and myeloid dendritic cells in HCC. Moreover, one research showed that paclitaxel suppressed HCC tumorigenesis through decreasing cell proliferation and accelerating programmed cell death by inhibiting the expression of YWHAZ (Liu et al., 2020). According to one study, YWHAZ gene, a positive regulator of p38 MAPK signaling pathway, induced vascular smooth muscle cell migration and intimal hyperplasia after vascular injury (Zhang et al., 2019b). These studies suggest that this gene could be an important biomarker in progression of HCC. In present study, this gene was enriched in shared biological processes in NASH, cirrhosis and HCC including negative regulation of programmed cell death, MAPK cascade, cytokine-mediated signaling pathway and regulation of mRNA stability. So far, the role of YWHAZ in cirrhosis and NASH has not been studied.

This gene may be a key biomarker in these diseases, and inhibition of this gene may prevent the progression of NASH and cirrhosis toward HCC.

In the present study, RAC1, HNRNPU, ITGAV and UBE2N genes were highlighted as key prognostic biomarkers with the most outdegree in NASH and HCC. Indeed, these candidate genes were enriched in several biological processes. The high expression of RhoA and RAC1 have been connected to development of various types of cancer (Yadav et al., 2020). RAC1 is important for different cellular functions, including proliferation, adhesion, motility, migration and metastasis of tumor cells (He et al., 2019, Toyama et al., 2010, Qi et al., 2019). The expression of RAC1 was significantly increased in NASH and HCC tissues, and the upregulation of RAC1 was associated with poor outcome for HCC patients (He et al., 2019, Li et al., 2018). RAC1 Inhibition may protect against progression of liver diseases (Soronen et al., 2016). RAC1 is negatively modulated by miR-576-5p, triggering a protective effect against NAFLD exacerbation (Caligiuri et al., 2016). Decreased RAC1 and RhoA activation leads to the suppression of HCC cell motility (Zhang et al., 2018). In the study of Xin Zhang et al. (Zhang et al., 2017b) RAC1 gene was enriched in VEGF signaling pathway, adherens junction and toll-like receptor (TLR) signaling pathway. RAC1 leads to hepatic lipoapoptosis. Lipoapoptosis is one of the important factors promoting the progression from hepatosteatosis to NASH (Huang et al., 2021). According to the study of Hongyi Yao et al. (Yao et al., 2015) RAC1 under-expression in endothelium diminished lung carcinoma-induced vascular permeability and decreased the metastasis to lungs. At present study, RAC1 was involved in multiple biological processes including cell-matrix adhesion, neutrophil degranulation, cell migration, VEGFR signaling, Wnt signaling, protein modification, Fc-gamma receptor signaling, ephrin receptor signaling, Rho signal transduction and receptor tyrosine kinase signaling. A recent study showed that activation of Wnt signaling contributes to the progression of human tumors and metastasis (Toyama et al., 2010). To date, the role of RAC1-related biological processes in the progression of NASH to HCC has not been studied. Therefore, *in vitro* and *in vivo* assessments need to be performed.

In recent study of Yi Liang et al. (Liang et al., 2021) HNRNPU promoted HCC progression by CDK2 overexpression. Furthermore, HNRNPU knockdown suppressed HCC cell proliferation and prompted cell cycle arrest. According to the study of Youli Du et al. (Du et al., 2021) the enhancement of HNRNPU transcription was associated with poor prognosis of HCC, and upregulated HNRNPU was a biomarker for identifying HCC. HNRNPU gene is an essential splicing regulator and over-expression of splicing factors may lead to aberrant splicing in tumors. The results of the study of Biao Zhang et al. (Zhang et al., 2021a) demonstrated that HNRNPU upregulation salvaged HCC cells from programmed cell death induced by c-Myc suppressors. In contrast, in one experimental study, HNRNPU underexpression enhanced inflammatory signaling in hepatocytes, thereby promoting liver injury and NASH progression. This research concluded that hepatic HNRNPU restricts the progression from steatosis to NASH (Xiong et al., 2020). In present study, HNRNPU was enriched in several biological processes including nucleic acid metabolism, RNA splicing, mRNA processing, transcription, mRNA stability, telomere lengthening, DNA biosynthesis, spindle organization, cell cycle, protein location in nucleus, positive regulation of ATPase activity and negative regulation of kinase activity. The role of HNRNPU-related biological processes in the progression of NASH to HCC has not been studied. Therefore, performing *in vitro* and *in vivo* assessments is necessary.

According to a new study (Wang et al., 2021) the expression of the ITGAV gene is increased in NASH. The findings of the research of Chun Lan Kang et al. (Kang et al., 2019) showed that upregulation of ITGAV gene was a potential landmark for metastasis or poor prognosis in HCC patients. Metastasis process is associated with cell migration and adhesion that are mostly integrin-dependent (Wang et al., 2016). A bioinformatics analysis showed that upregulated ITGAV gene was detected as a hub gene in HCC (He et al., 2017). In one in vitro study, lncRNA AY promoted HCC metastasis via ITGAV overexpression (Kang et al., 2019). In present study, ITGAV was enriched in multiple biological processes including negative regulation of apoptotic signaling pathway, cell-cell adhesion, cell-matrix adhesion, integrin signaling, endocytosis, LDL receptor activity, neutrophil degranulation, antigen processing and presentation, cell proliferation, receptor tyrosine kinase signaling, VEGFR signaling and lipid transport. The role of ITGAV-related biological processes in the progression of NASH to HCC has not been studied. Therefore, performing more researches is essential.

According to a systems biology analysis, UBE2N gene was detected as hub gene in NASH (Karbalaei et al., 2017). Recent studies demonstrate the involvement of UBE2N in progression of melanoma, HCC, breast, prostate, lymphoma, and ovarian cancer (Singh and Sharma, 2020, Wang et al., 2019b). In contrast, in one *in vitro* study, the enhancement of UBE2N transcription considerably repressed tumor cell proliferation of HCC (Zhang et al., 2017a). In present study, UBE2N was implicated in several biological processes including selective autophagy, cytokine-mediated signaling, antigen receptor-mediated signaling, DNA metabolism, histone ubiquitination, protein ubiquitination, protein modification, histone modification and positive regulation of transferase activity. The role of UBE2N-related biological processes in the progression of NASH to HCC has not been studied. Therefore, performing further studies is needed.

In the present study, ITGB1, PRKCI, PPIA and RAB7A genes were highlighted as prognostic biomarkers with the most outdegree in HCC. Indeed, these candidate genes were enriched in several biological processes. Integrins play important roles in cell adhesion, anti-apoptosis, migration and proliferation (Lee et al., 2011). Extracellular matrix (ECM) remodeling has been detected in HCC microenvironment (Xie et al., 2021). ITGB1, an extracellular matrix (ECM) receptor, is a well-known oncogene that plays a critical role in growth and metastasis of HCC (Zhang et al., 2019a). The recent study of Jinghe Xie et al. (Xie et al., 2021) showed that ITGB1 and YWHAZ were meaningfully overexpressed in HCC tissue in comparison with adjacent normal tissue and the inhibition of ITGB1 expression by small interfering RNA (siRNA) resulted in the downregulated expression of YWHAZ in HCC cells. ITGB1 knockdown obviously disturbed the violent behavior of HCC tumor cells and delayed cell cycle development. In addition, high expression of ITGB1 gene reversed the suppressive effects of miR-3653 on the growth, metastasis and epithelial-mesenchymal transition (EMT) of HCCLM3 cells (Zhang et al., 2019a). ITGB1 gene was closely associated with angiogenesis and neovascularization (Qi et al., 2020, Lu et al., 2018). In one study, the increase of ITGB1 transcription was involved in ovarian carcinogenesis induced by inflammatory cytokines, and targeting ITGB1 was an efficient therapeutic strategy (Yang et al., 2014). According to the study of Teijo Pellinen et al. (Pellinen et al., 2018) in prostate cancer, ITGB1-associated overexpression of Caveolin-1 switches TGFβ signaling from tumor-inhibitor to oncogenic. In present study, ITGB1 was involved in several biological processes including integrin signaling, angiogenesis, regulation of apoptotic process, cell migration, cell-matrix adhesion, cytokine signaling, hydrolase activity, TGF beta receptor signaling, response to LDL, GTPase activity, as well as adhesion, rolling and differentiation of lymphocyte. The participation of ITGB1-related biological processes in HCC progression need to be assessed.

According to the study of Nada M.K. Mabrouk et al. (Mabrouk et al., 2020) upregulation of PRKCI gene was observed in HCC patients. PRKCI, a human oncogene, plays a prominent role in cell proliferation, differentiation, and carcinogenesis (Qu et al., 2016). Liujing Qu et al. (Qu et al., 2016) showed that PRKCI overexpression in U2OS cells impaired functional autophagy. PRKCI downregulation promotes autophagy (Moscat and Diaz-Meco, 2020). Moreover, PRKCI enhances immune suppression in ovarian cancer (Sarkar et al., 2017). According to the study of Itaru Hashimoto et al. (Hashimoto et al., 2019) after gastrectomy, high expression of PRKCI in cancerous tissue might be a valuable prognostic factor in gastric cancer. In present study, PRKCI was enriched in several biological processes including regulation of apoptotic process, cell-cell junction, insulin stimulus, Notch signaling, proliferation, protein modification, protein localization, signal transduction, cytoskeleton organization and protein phosphorylation. The participation of PRKCI-related biological processes in developing HCC need to be assessed.

According to the study of Yuwei Gu et al. (Gu et al., 2021) there is a strong correlation between the expression levels of PPIA and the immune cell infiltration or the expression of chemokines. In contrast, Hye Ri Ahn et al. (Ahn et al., 2021) showed that PPIA expression was stable in liver diseases. High expression of PPIA transcription was associated with reduced overall survival in glioma, acute myeloid leukemia, lung adenocarcinoma, skin cutaneous melanoma and HCC (Wang et al., 2019a). In present study, PPIA was implicated in several biological processes including downregulation of intrinsic apoptotic signaling pathway in response to oxidative stress, cytokine signaling, DNA biosynthesis, MAPK cascade, neutrophil degranulation, oxidative stress, protein modification, protein phosphorylation and signal transduction. The participation of PPIA-related biological processes in progressing HCC need to be assessed.

Jiming Xie et al. (Xie et al., 2019) showed that RAB7A suppressed the apoptosis and promoted proliferation and invasion of breast cancer cells. The results of the study of Ting-ting Liu et al. (Liu and Liu, 2020) indicated that RAB7A may serve as a prognostic biomarker in colorectal cancer as well as a new target to treat colorectal cancer. In present study, RAB7A was enriched in various biological processes including antigen processing and presentation, negative regulation of programmed cell death, EGFR signaling, neutrophil degranulation, selective autophagy, protein localization and phagosome acidification. The participation of RAB7A-related biological processes in exacerbating HCC need to be assessed.

In the present study, GSK3B, DAB2, PSME4 and PPP3CA genes were highlighted as prognostic candidate biomarkers with the most outdegree in NASH. Ya-Ling Yang et al. (Yang et al., 2020) showed that GSK3β gene could serve as possible therapeutic target to improve NASH. One study showed that GSK3B downregulation promoted programmed cell death in breast cancer (Xu et al., 2021). Guoxu Fang et al. (Fang et al., 2019) showed that suppression of GSK-3β activity reduced glucose consumption and adenosine triphosphate (ATP) level in HCC cells. Therefore, inhibition of GSK-3β lessened viability and metastasis of HCC cells. In present study, GSK3β was involved in multiple biological processes including negative regulation of non-apoptotic programmed cell death, regulation of autophagy, cell-matrix adhesion, glucan metabolism, glycogen metabolism, glycogen biosynthesis, protein modification, response to cytokine, response to insulin, response to stress, Wnt signaling, ubiquitination, tyrosine kinase signaling and morphogenesis. The contribution of GSK3β-related biological processes in developing NASH need to be assessed.

Bioinformatics analysis of an animal model of diet-induced nonalcoholic fatty liver disease demonstrated that DAB2 was a poor prognostic gene for HCC (Hong et al., 2022). Contrary to the present study, some studies showed that DAB2 may function as a tumor suppressor and its lower expression has been found in several cancers (Hocevar, 2019, Wang et al., 2020). Chunhui Sun et al. (Sun et al., 2018) showed that miR-106b may elevate HCC cell proliferation and migration by targeting DAB2. Moreover, Liqin Du et al. (Du et al., 2014) revealed that miR-93 oncogenic function is mediated by decreasing DAB2 transcription. In present study, DAB2 was implicated in the biological processes including androgen receptor signaling, negative regulation of programmed cell death, cell migration, endocytosis, integrin signaling, ubiquitination and Wnt signaling. The contribution of DAB2-related biological processes in developing NASH need to be explored.

The role of PPP3CA and PSME4 genes in NASH has not yet been identified. Sijia Ge et al. (Ge et al., 2022) revealed that the overexpression of PSME4 promoted the proliferation of HCC cells through the mTOR signaling pathway and PSME4 is a prognostic biomarker for the initial diagnosis of HCC. Yongdong Guo et al. (Guo et al., 2021) showed that PSME4 gene could serve as a worthy diagnostic biomarker for gastric cancer. The results of one study indicated a possible correlation between PPP3CA overexpression and the pathogenesis of multiple myeloma (Imai et al., 2016). In present study, PSME4 gene was enriched in multiple biological processes including antigen processing, cytokine-mediated signaling, Fc-epsilon receptor signaling, interleukin-1 signaling, Wnt signaling, T cell receptor signaling, protein ubiquitination, protein modification and mRNA catabolism. Moreover, PPP3CA gene was enriched in various biological processes including calcium modulating, cell adhesion, cell migration, Fc-epsilon receptor signaling, protein modification, Wnt signaling, transcription and wound healing. The contribution of PSME4 and PPP3CA-related biological processes in developing NASH need to be studied.

In the present study, CCR7, ENG, RHOQ, and HDAC7 genes were highlighted as prognostic candidate biomarkers with the most outdegree in liver cirrhosis. The results of the study of Wufeng Fan et al. (Fan and Ye, 2018) revealed that CCR7 was identified as the hub gene in cirrhosis-originating HCC. In present study, CCR7 was enriched in several biological processes including cell adhesion, cell motility, cytokine production, dendritic cell chemotaxis, dendritic cell differentiation, dendritic cell migration, interleukin-1 production, neutrophil migration, neutrophil chemotaxis, T cell activation, T cell polarity, response to cytokine, interferon-gamma production, immunological synapse, leukocyte apoptotic process, MAPK cascade, ERK1 and ERK2 cascade. The contribution of CCR7-related biological processes in development of liver cirrhosis need to be discoverd.

The expression of endoglin (ENG) isoforms is differentially enhanced in chronic and acute liver damages (Alsamman et al., 2018). Moreover, ENG is involved in arterialization and sinusoidal capillarization process in HCC (Kasprzak and Adamek, 2018). According to one study, serum ENG is significantly elevated in liver cirrhotic patients compared with healthy controls (Yagmur et al., 2007). In contrast, some researches have shown that ENG downregulation in HCC tissue and its serum levels potentially serve as a poor prognostic marker in HCC patients and downregulation of ENG is associated with the epithelial–mesenchymal transition (EMT) promotion (Eun et al., 2020). Muhammad Alsamman et al. (Alsamman et al., 2018) showed that ENG has protective role against fibrotic injury, likely through modulation of TGF-β signaling. In an experimental study, Ivone Cristina Igreja Sá et al. (Igreja Sá et al., 2020) showed that ENG might be a biomarker of NASH development and the presence of high levels of ENG might contribute in NASH exacerbation. In present study, ENG was enriched in multiple biological processes including cell migration, cell proliferation, mesenchymal cell differentiation, smooth muscle cell differentiation, response to TGF beta stimulus, morphogenesis and regulation of apoptosis. The contribution of ENG-related biological processes in development of liver cirrhosis need to be identified.

The role of RHOQ gene in liver cirrhosis has not yet been identified. High expression of RHOQ gene in colon adenocarcinoma and gastric cancer was with poor prognosis (Hu et al., 2019, Zhang et al., 2016). In present study, RHOQ was enriched in several biological processes including cytoskeleton organization, filopodium assembly, glucose import, GTP metabolism, protein localization, signal transduction and tyrosine kinase signaling. The contribution of RHOQ-related biological processes in progression of liver cirrhosis need to be explored.

Mohammed Nihal Hasan et al. (Hasan et al., 2019) showed that HDAC7 Knockdown led to cycle cell arrest in various cancer cells. In the research of Rajeswara Rao Pannem et al. (Pannem et al., 2014) binding HDAC7 to the promoter of HGF gene led to under-expression of HGF, the gene that restrains liver fibrogenesis, in hepatic stellate cells. In a recent study, MiR-489 repressed tumor proliferation and invasion in colorectal cancer by targeting HDAC7 (Gao et al., 2018). In present study, HDAC7 was involved in the biological processes including cell migration, interleukin-2 production, cytokine production, NFKB signaling and osteoblast differentiation and protein deacetylation. The contribution of HDAC7-related biological processes in development of liver cirrhosis need to be assessed.

5. **Conclusion:**

In present study, for the first time, novel prognostic genes and key biological processes were uncovered in NASH, cirrhosis and HCC based on overall survival and using systems biology approach. Hence, performing *in vitro* and *in vivo* researches to verify the results is necessary.

**Table1-Biological processes related to YWHAZ (p<0.05)**

| | | p value |
|---|---|---|
| NASH | cytokine-mediated signaling pathway (GO:0019221) | 0.0002 |
| | protein phosphorylation (GO:0006468) | 0.0008 |
| | negative regulation of programmed cell death (GO:0043069) | 0.003 |
| | cellular protein modification process (GO:0006464) | 0.003 |
| | cellular component assembly (GO:0022607) | 0.003 |
| | regulation of MAPK cascade (GO:0043408) | 0.004 |
| | regulation of mRNA stability (GO:0043488) | 0.01 |
| | regulation of mRNA catabolic process (GO:0061013) | 0.02 |
| | cellular response to cytokine stimulus (GO:0071345) | 0.03 |
| | positive regulation of mitochondrial outer membrane permeabilization involved in apoptotic signaling pathway (GO:1901030) | 0.04 |
| HCC | regulation of apoptotic process (GO:0042981) | 0.000002 |
| | cellular protein modification process (GO:0006464) | 0.000005 |
| | negative regulation of programmed cell death (GO:0043069) | 0.00007 |
| | cytokine-mediated signaling pathway (GO:0019221) | 0.0001 |
| | regulation of mRNA stability (GO:0043488) | 0.0003 |
| | negative regulation of apoptotic process (GO:0043066) | 0.0004 |
| | regulation of mRNA catabolic process (GO:0061013) | 0.002 |
| | cellular component assembly (GO:0022607) | 0.006 |
| | positive regulation of protein localization to membrane (GO:1905477) | 0.01 |
| | regulation of MAPK cascade (GO:0043408) | 0.03 |
| | protein phosphorylation (GO:0006468) | 0.03 |
| Cirrhosis | cytokine-mediated signaling pathway (GO:0019221) | 1E-08 |
| | cellular response to cytokine stimulus (GO:0071345) | 2E-07 |
| | regulation of ERK1 and ERK2 cascade (GO:0070372) | 0.000006 |
| | regulation of apoptotic process (GO:0042981) | 0.000009 |
| | cellular component assembly (GO:0022607) | 0.0001 |
| | regulation of MAPK cascade (GO:0043408) | 0.0003 |
| | negative regulation of programmed cell death (GO:0043069) | 0.0008 |
| | negative regulation of apoptotic process (GO:0043066) | 0.002 |
| | establishment of Golgi localization (GO:0051683) | 0.01 |
| | Golgi inheritance (GO:0048313) | 0.02 |

| | |
|---|---|
| regulation of mRNA stability (GO:0043488) | 0.03 |
| positive regulation of intracellular protein transport (GO:0090316) | 0.03 |
| regulation of protein insertion into mitochondrial membrane involved in apoptotic signaling pathway (GO:1900739) | 0.04 |

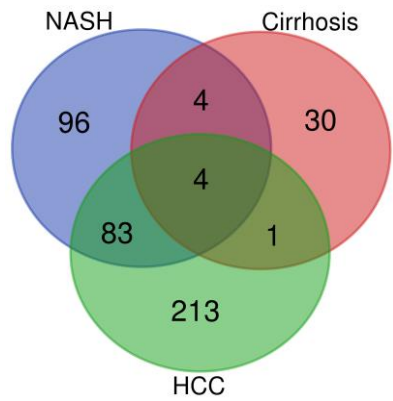

Figure1: Identification of the shared and specific overexpressed DEGs in NASH, Cirrhosis and HCC (based on inclusion criteria) via a Venn diagram. The blue circle indicates NASH, the green circle indicates HCC, and the red circle indicates cirrhosis.

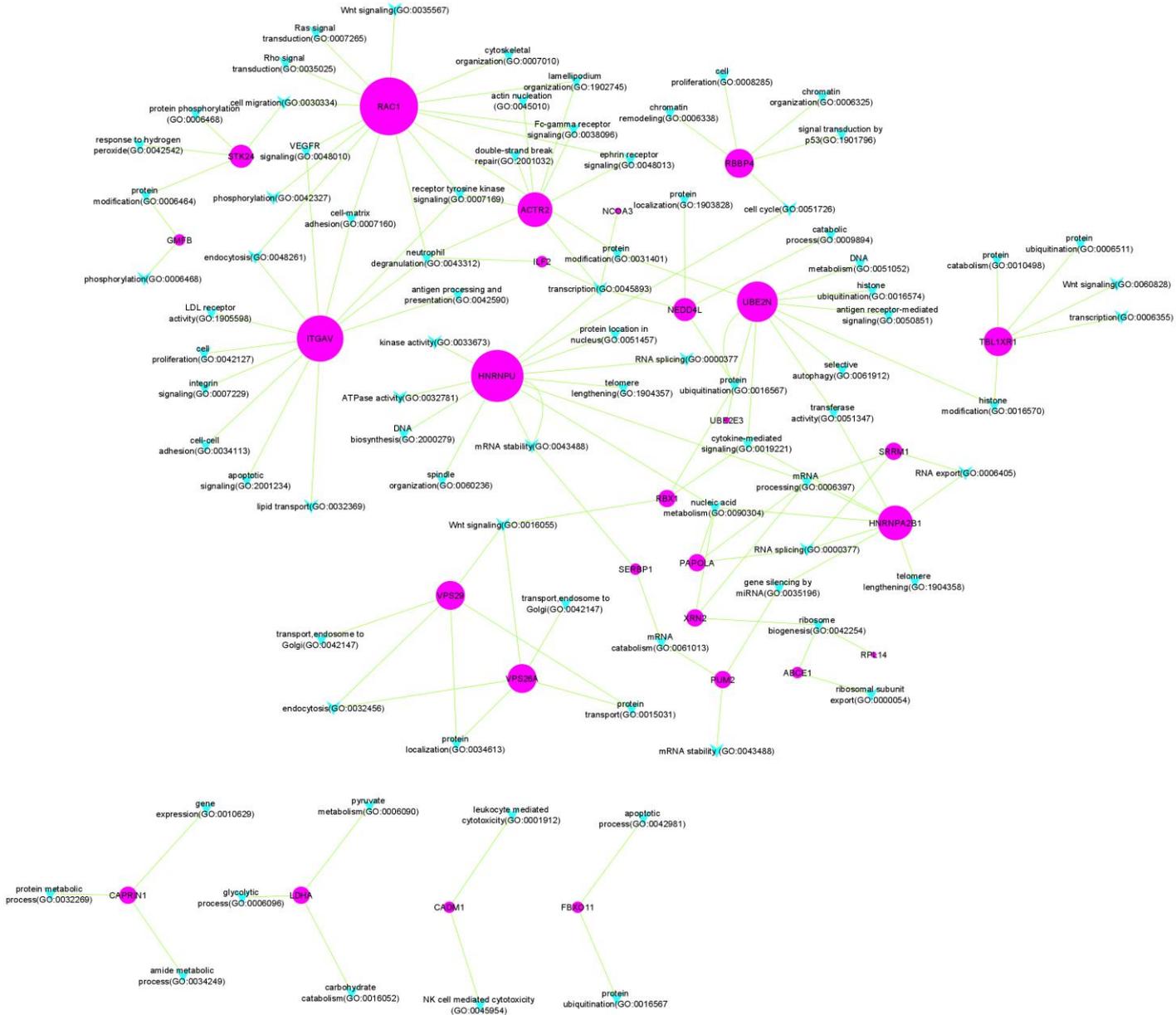

**Figure2:** The gene-biological process network. The pink nodes represent shared prognostic genes between HCC and NASH, and the blue nodes represent shared biological processes between HCC and NASH. The edges represent the interactions between them and size of pink nodes is proportional to outdegree of the genes.

**Figure3: The gene-biological process network in HCC. The pink nodes represent specific prognostic genes in HCC, and the blue nodes represent biological processes associated with specific prognostic genes in HCC. The edges represent the interactions between them and size of pink nodes is proportional to outdegree of the genes.**

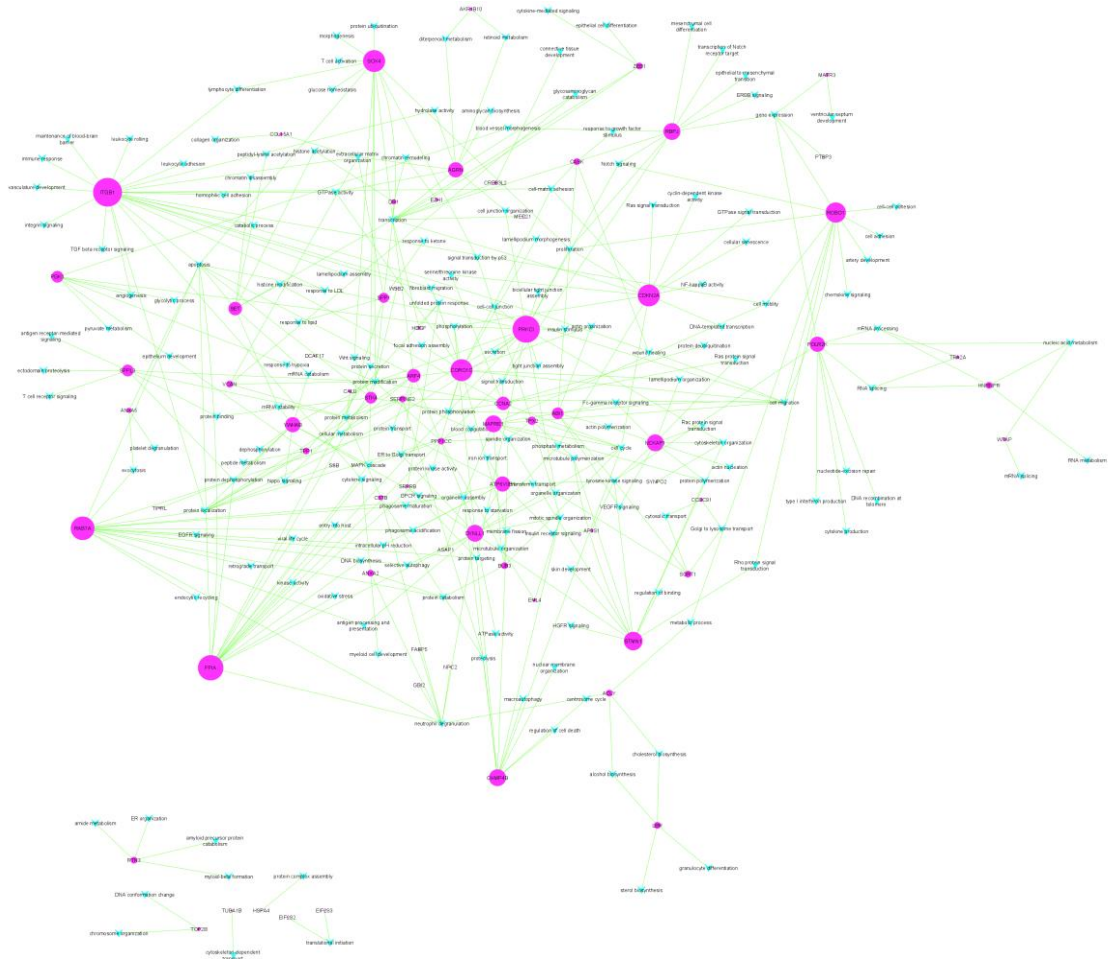

**Figure4: The gene-biological process network in NASH. The pink nodes represent specific prognostic genes in NASH, and the blue nodes represent biological processes associated with specific prognostic genes in NASH. The edges represent the interactions between them and size of pink nodes is proportional to outdegree of the genes.**

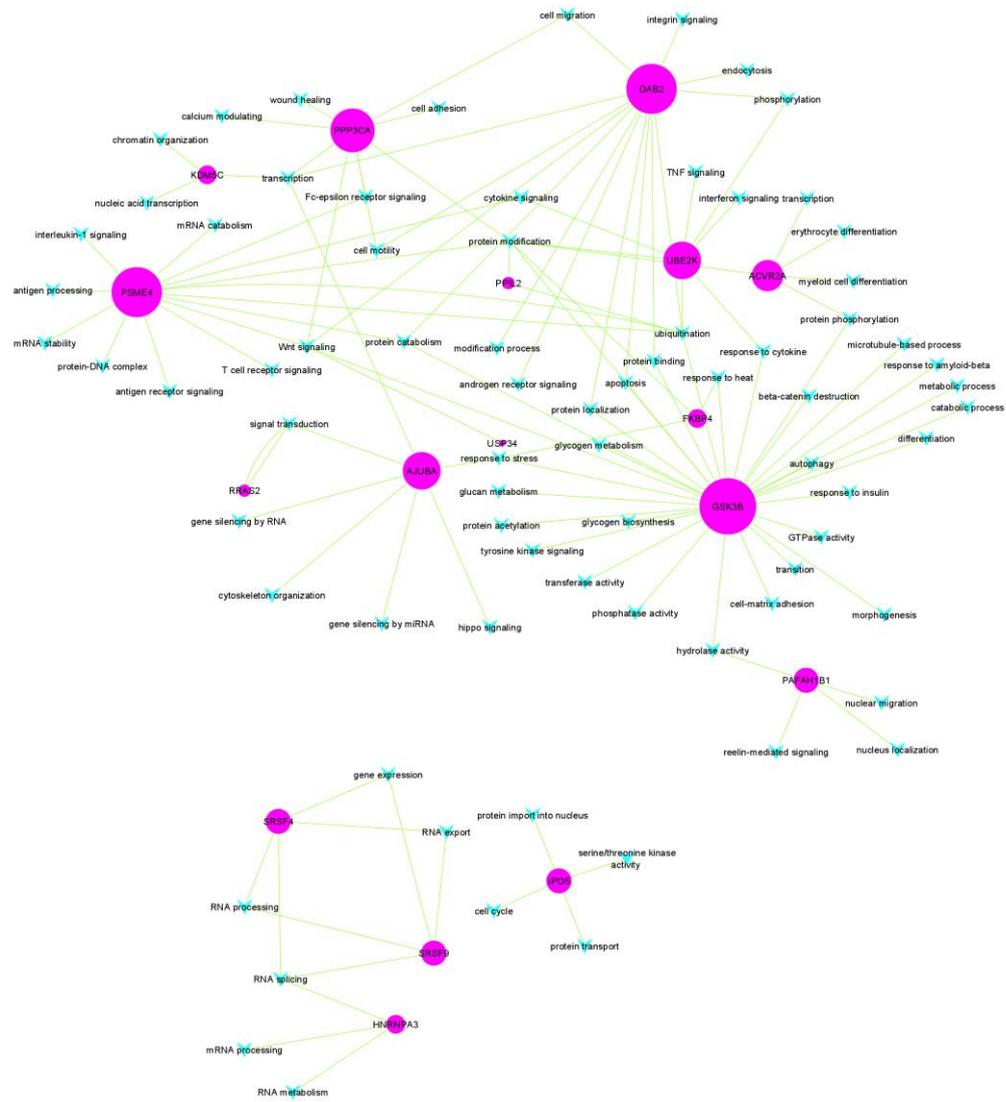

**Figure 5: The gene-biological process network in cirrhosis. The pink nodes represent specific prognostic genes in cirrhosis, and the blue nodes represent biological processes associated with specific prognostic genes in cirrhosis. The edges represent the interactions between them and size of pink nodes is proportional to outdegree of the genes.**